\documentclass[reprint,superscriptaddress,longbibliography,amsmath,amssymb,aps,prl]{revtex4-1}

\usepackage{graphicx}
\usepackage{dcolumn}% Align table columns on decimal point
\usepackage{wasysym}
\usepackage{siunitx, textcomp}
\usepackage{makecell}
\usepackage[rightcaption]{sidecap}
\sidecaptionvpos{figure}{t}
\usepackage{booktabs}
\begin{document}

%\title{Radiation Efficiency and Absorption Losses of Implantable Bioelectronics}%

%\title{Strategies to Improve Electromagnetic Efficiency of \\ Wireless Implantable Bioelectronics}%

%\title{Strategies to Improve Electromagnetic Efficiency of Implantable RF Bioelectronics}% DN: we can improve the title to be terrer understandable by CELL readers, e.g.:

\title{Physical Insights into Electromagnetic Efficiency \\ of Wireless Implantable Bioelectronics}%

\author{Mingxiang Gao}%
\affiliation{Microwaves and Antennas Group, Ecole Polytechnique Federale de Lausanne, Lausanne, Switzerland}%

\author{Denys Nikolayev}%
\affiliation{Univ Rennes, CNRS, IETR – UMR 6164, FR-35000 Rennes, France}%

\author{Zvonimir Sipus}%
\affiliation{Faculty of Electrical Engineering and Computing, University of Zagreb, Zagreb, Croatia}%

\author{Anja K. Skrivervik}%
\email{anja.skrivervik@epfl.ch}%
\affiliation{Microwaves and Antennas Group, Ecole Polytechnique Federale de Lausanne, Lausanne, Switzerland}%

%\date{\today}

\begin{abstract} % abstract

\textbf{Background} Autonomous implantable bioelectronics rely on wireless connectivity, necessitating highly efficient electromagnetic (EM) radiation systems. However, limitations in power, safety, and data transmission currently impede the advancement of innovative wireless medical devices, such as tetherless neural interfaces, electroceuticals, and surgical microrobots. To overcome these challenges and ensure sufficient link and power budgets for wireless implantable systems, this study explores the mechanisms behind EM radiation and losses, offering strategies to enhance radiation efficiency in wireless implantable bioelectronics.

\textbf{Methods} Using analytical modeling, the EM waves emitted by the implant are expanded as a series of spherical harmonics, enabling a detailed analysis of the radiation mechanisms. This framework is then extended to approximate absorption losses caused by the lossy and dispersive properties of tissues through derived analytical expressions. For the first time in the literature, the radiation efficiency and in-body path loss are quantified and compared in terms of three primary loss mechanisms.

\textbf{Results} The impact of various parameters on the EM efficiency of implantable devices is analyzed and quantified, including operating frequency, implant size, body--air interface curvature, and implantation location. Additionally, a rapid estimation technique is introduced to determine the optimal operating frequency for specific scenarios, along with a set of design principles aimed at improving radiation performance.

\textbf{Conclusions} The design strategies derived in this work---validated through numerical and experimental demonstrations on realistic implants---reveal a potential improvement in implant radiation efficiency or gain by a factor of five to ten, leading to a corresponding increase in overall link efficiency compared to conventional designs. Due to the simplicity of the derived closed-form expressions, these design principles are particularly useful during the early stages of wireless implant development.
\end{abstract}

\maketitle

%%%%%%%%%%%%%%%%%%%%%%% INTRODUCTION %%%%%%%%%%%%%%%%%%%%%%%%%%%%%%%%
\section*{Introduction}
\label{sec:intro}

Implantable bioelectronics are increasingly being used to interact with biological systems, unlocking remarkable opportunities in diagnostics, treatments, and fundamental scientific exploration \cite{song2019wearable, veleticImplantsSensingCapabilities2022, famm_drug_2013, won2020emerging, joo2021soft, won2021wireless}. To achieve long-term operation of implantable devices and improve their clinical capabilities, efficient and safe wireless data transmission and power transfer through body tissues remain a major challenge \cite{xie2020flexible, agrawal2017conformal, kim_wireless_2012, PhysRevApplied.8.014031, yang2019enhancing}. Bioelectronics typically rely on radio-frequency (RF) electromagnetic (EM) waves for through-body communications and powering~\cite{won2021wireless}, while other approaches, such as ultrasonic methods, have been proposed as well \cite{piechWirelessMillimetrescaleImplantable2020}. To compensate for the lossy, heterogeneous, and non-deterministic nature of biological tissues, existing implantable RF devices require wireless transmission at relatively high power levels \cite{Agarwal_2017}. Compared to increasing the transmitted power, operating at optimal radiation conditions is clearly a better option due to the constraints on safety and power budget \cite{nikolayev2019optimal}. Reducing the power consumption without functional compromises could improve the miniaturization and lifetime of wireless implantable devices \cite{jungInjectableBiomedicalDevices2020}. In this context, a comprehensive understanding of the EM radiation mechanism of implantable bioelectronics will allow for lower tissue absorption losses and more precise EM field control (e.g., bio-adaptive in-body radiation sources and on-body wavefront manipulation devices), thereby enabling close-to-optimal wireless performance.

A typical wireless system for an implantable RF device establishes a wireless link between the implant and an external unit \cite{Yoo_2021}. Owing to the reciprocity of antennas, as in general wireless systems, the wireless link efficiency between the implantable antenna and the external antenna remains constant whether the implant is a transmitter or a receiver. Although advances in RF technologies have enabled the widespread deployment of wireless systems, establishing wireless links for implantable devices is still challenging. The fundamental challenge is the significant EM attenuation due to the high loss nature of biological tissues, i.e., absorption losses that occur within the host body \cite{Chow_2013}. For the implant itself, a key indicator to quantify EM absorption is the radiation efficiency of the antenna, as the latter is directly linked to the antenna gain in a specific direction, and is severely reduced by in-body losses \cite{Khan_2022}. This results in a very tight or even insufficient link budget, such as a reduction in received power and a deterioration of signal integrity. Therefore, investigating the EM absorption of implantable bioelectronics to achieve high power efficiency has become a core concern in engineering design \cite{Li_2021}.

 EM radiation mechanisms of arbitrary EM sources in free space have been extensively studied since the 1940s \cite{wheeler1947fundamental, chu_physical_1948, harrington1958gain, gustafsson_physical_2007}. However, these results cannot be directly translated into considerations for sources in lossy heterogeneous biological tissues. In this case, the radiation efficiency of the implant cannot be decoupled from the surrounding medium and is strongly affected by the near-field coupling with the medium losses, wave scattering, attenuation, and effects on the body interfaces \cite{Chow_2013, Kiourti2014}. Moreover, the miniature dimensions of implantable bioelectronics require implants much smaller than the wavelength, which is another factor limiting the radiation performance \cite{Khalifa_2021}. In terms of modeling, the use of anatomically-realistic body models enables accurate analysis of the radiation characteristics, specific to the implant location and application \cite{nikolayev_electromagnetic_2018, soaresWirelessPoweringEfficiency2022a}. However, they are time-consuming and thus not suitable for pre-design assessment of the feasibility of wireless links. In contrast, simplified body models use multiple layers of media to represent biological tissues and air in general terms~\cite{gao_analytical_2024}. These models are not only suitable for the rough design of implants but also for analytical modeling to gain insight into radiation mechanisms. Making use of layered spherical body models, resonance characteristics and radiation signature of implanted antennas were first assessed \cite{kim2004implanted}, followed by the determination of the optimal radiating conditions for finite-sized capsule implants \cite{nikolayev2019optimal}. Based on the layered planar body model, optimal frequency and analytical bounds for wireless power transfer to an implanted device were obtained \cite{poon_optimal_2010, kim_midfield_2013}.
Using the spherical harmonic expansion of EM waves \cite{bosiljevac2015propagation}, the fundamental limitations of implanted antennas are evaluated, where the near-field loss for different harmonic modes is expressed analytically for the first time \cite{skrivervik2019fundamental}.
Besides investigating the implant itself, techniques to mitigate the loss due to wave-impedance mismatch on a body--air interface were investigated using wearable diffractive patterns and metasurfaces \cite{yangAntireflectionWavefrontManipulation2020, yang2019enhancing}.

In this work, we first investigate the EM radiation mechanism of wireless implantable devices by analytical modeling and closed-form expressions to assess radiation efficiency. Building on this foundation, we introduce, for the first time, a rapid estimation technique for determining the optimal frequency and a set of design principles aimed at maximizing implant radiation efficiency. These design rules and procedures are validated through several numerical and experimental demonstrations of realistic implants, revealing a 5- to 10-fold improvement in wireless link efficiency after optimization.

\section*{Methods}

\subsection*{Analytical Modeling of Implanted EM Source}
To analyze EM radiation of body-implanted bioelectronics, analytical modeling of the implant can facilitate a comprehensive and effective understanding despite being only an approximation compared to the real body environment \cite{Chow_2013}. In this work, a stratified spherical body phantom of variable radius ${r_{{\rm{body}}}}$ (represents the curvature radius of the body-air interface) and of complex permittivity $\varepsilon_{\rm{body}} (\omega) = \varepsilon_{\rm{body}}' - i\varepsilon_{\rm{body}} ''$ ($\omega  = 2\pi f$ is the angular frequency and $f$ is the operating frequency) is used to model the host body of the implant, as depicted in Fig.~\ref{fig1}. 

\begin{figure}[!b]% (b,c) -> el. mag sources + caption
\centering
\includegraphics[scale=0.37]{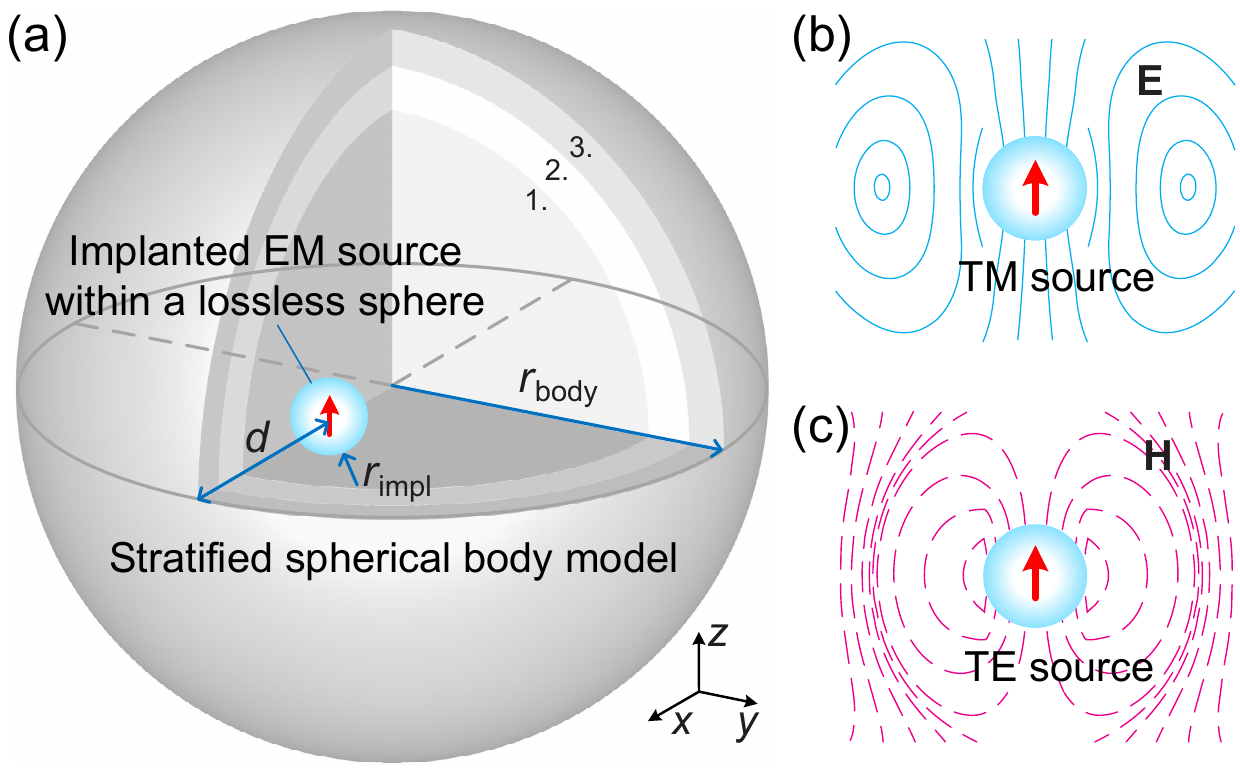}
\caption{\label{fig1} (a) View of the formulated spherical body model: a stratified spherical body model of dispersive body tissue with an implanted source surrounded by a small lossless sphere. For a typical body phantom, 1 is the muscle layer, 2 is the fat layer and 3 is the skin layer. The distribution of (b) electric fields \textbf{E} for a TM source and (c) magnetic fields \textbf{H} for a TE source.}
\end{figure} 

Considering the operating frequencies and related power levels, properties of body tissue are assumed isotropic, nonmagnetic (permeability $\mu_{\rm{body}}  = {\mu _0}$), and linear, i.e., E-field amplitude does not affect $\varepsilon$ \cite{nikolayev2019optimal, Schmidt_2012}. To represent the dispersion of lossy body tissues, we use the four-region Cole--Cole model, whose specific parameters have been defined in \cite{gabriel1996dielectric} based on the experimental data. The implanted source, an elementary electric dipole source (i.e., the source exciting the transverse magnetic modes, abbreviated as TM source) or magnetic dipole source (i.e., the source exciting the transverse electric modes, abbreviated as TE source), is surrounded by a small lossless sphere of radius ${r_{{\rm{impl}}}}$ (represents the encapsulation size of an antenna fitted in an implantable device). The dipole moment is oriented in the $z$-direction. For the model under study, the implanted source can be offset at a distance ${r_{{\rm{feed}}}}$ from the center of the model to account for the effects of implantation depth, i.e. $d$, and the curvature radius of the body--air interface, i.e. ${r_{{\rm{body}}}}$.

In this study, we assume that the meticulous task of antenna matching has been adeptly executed by the antenna engineer, as evidenced by the existence of implantable antennas impervious to detuning \cite{nikolayevImmunetodetuningWirelessInbody2019}. Hence, the ohmic loss is not considered in the modeling because we regard the antenna mismatch losses as insignificant in comparison to the other losses investigated in this study.

\subsection*{Spherical Harmonic Analysis of~Electromagnetic~Waves in Tissues}
The derivation of the radiation efficiency and loss components relies on the spherical wave expansion (SWE), in which EM fields $\mathbf{E}$ and $\mathbf{H}$ in spherical geometry (with zero free-charge density) are decomposed into vector spherical harmonics \cite{stratton1941electromagnetic, harrington1961time, chew1995waves} as follows
\begin{equation}
\left\{ {\begin{array}{*{20}{c}}
  {\mathbf{E}} \\ 
  { - i\zeta {\mathbf{H}}} 
\end{array}} \right\} = \sum\limits_{n,m} {{a_{mn}}\left\{ {\begin{array}{*{20}{c}}
  {{{\mathbf{M}}_{mn}}} \\ 
  {{{\mathbf{N}}_{mn}}} 
\end{array}} \right\}}  + {b_{mn}}\left\{ {\begin{array}{*{20}{c}}
  {{{\mathbf{N}}_{mn}}} \\ 
  {{{\mathbf{M}}_{mn}}} 
\end{array}} \right\}.
\end{equation}
\noindent Specifically, the vector spherical harmonics ${\mathbf{M}}_{mn}$ and ${\mathbf{N}}_{mn}$ are written as
\begin{equation}
{{\mathbf{M}}_{mn}} = \nabla  \times {\mathbf{r}}{\psi _{mn}},
{{\mathbf{N}}_{mn}} = \frac{1}{k}\nabla  \times \nabla  \times {\mathbf{r}}{\psi _{mn}},
\end{equation}
\begin{equation}
{\psi _{mn}} = \frac{1}{{kr}}{\hat Z_n}(kr)P_n^m(\cos \theta ){e^{im\varphi }},
\end{equation}
\noindent where $\mathbf{r}$ is the radius vector from the origin (the radial distance is $r$), $k = \omega \sqrt {{\mu _0}\varepsilon }  = k' - ik''$ is the wave number, ${\psi _{mn}}$ is the scalar solution to the Helmholtz equation in the spherical coordinate system, ${\hat Z_n}$ denotes the Schelkunoff type spherical Bessel or Hankel functions of order $n$, and $P_n^m$ are the associated Legendre polynomials of degree $n$ and order $m$. Using the orthogonality properties of spherical harmonics, the radiated power through a concentric spherical surface at any radius can be expressed analytically. The main steps of the SWE method are presented in \cite{bosiljevac2015propagation}, and the spherical geometry can be made stratified through a mode matching technique \cite{Sanford_1994, chew1995waves}.

By numerical implementation, the EM field at any point, as well as the radiated power through a concentric sphere, can be obtained. Thus, we calculate the total radiation efficiency for implanted sources as ${\eta_{\rm{total}}}={P_{\rm{air}}} / {P_{\rm{impl}}} $, where ${P_{\rm{air}}}$ is the total radiated power reaching free space 
(air), and ${P_{\rm{impl}}}$ is the power radiated by the implant (i.e., the power entering the lossy medium). For a typical spherical model with an implanted off-center source ($r_{{\rm{feed}}}>0$), the spherical body phantom and the offset sphere surrounding the source can be connected using addition theorems \cite{stein1961addition}. By the equivalence principle, if ${r_{{\rm{impl}}}}\ll \lambda _{\rm{body}}$ ($\lambda _{\rm{body}}=2\pi /k'_{\rm{body}}$ is the wavelength in the body tissue), a simpler method is to replace the lossless sphere with the same body medium, and the value at the same lossless sphere boundary is calculated and used for normalization.

\begin{figure}[!t]
\centering
\includegraphics[scale=0.6]{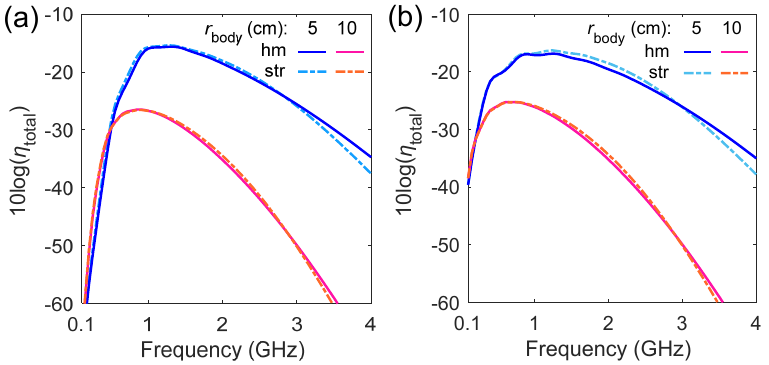}
\caption{\label{fig2} Frequency dependency of ${\eta_{\rm{total}}}$: (a) a TM source and (b) a TE source ($r_{\rm impl} = 5$ mm) implanted at the center of a spherical body phantom. The body phantoms applied include: a homogeneous (hm) muscle phantom with a radius ${r_{{\rm{body}}}}$ and a three-layer stratified (str) phantom with the additional 3-mm fat layer and 2-mm skin layer outside the muscle phantom.}
\end{figure}

To study the radiation properties of the implanted source, a simplified homogeneous body model can provide useful results in most cases as an initial and quick assessment \cite{poon_optimal_2010, skrivervik2019fundamental}. For instance, with the introduction of relatively thin layers of fat and skin (see Fig.~\ref{fig2}), the radiation efficiency in the homogeneous muscle phantom is generally similar to that of the stratified phantom regardless of source types or implantation depths. This suggests that these thin body layers have no significant effect on ${\eta_{\rm{total}}}$ compared to the dominant body tissue surrounding the implant. For more complex implantation locations, such as the presence of surrounding organs or thick body layers, tissue-equivalent media are commonly used for engineering estimates prior to specific numerical analysis \cite{FCCguidelines}.

\subsection*{Closed-Form Approximation of Radiation Efficiency}
The spherical body model with a centered source provides a concise analytical solution for analyzing electromagnetic radiation from the implant to free space with a constant implantation depth in all directions. Due to the spherical symmetry of this model, it can achieve the maximum radiation efficiency for a constant implantation depth. In addition, since the radiation pattern of the implanted source is not disturbed by the surrounding body phantom, the obtained radiation efficiency can also be directly used to characterize the loss of wireless link or antenna gain caused by the body phantom in any direction. 

For a homogeneous spherical body model with a center source ($r_{{\rm{feed}}}=0$), the dominant spherical mode becomes the lowest spherical mode with $n=1$, which contributes to the lowest near-field loss and reflection loss \cite{skrivervik2019fundamental}. On this basis, the radiation efficiency of the implanted source can be analytically approximated by an upper bound and decomposed into three main loss contributions: 1)~the efficiency due to the loss in the coupling of the reactive near field and the lossy biological tissue ${\eta_{{\rm{near~field}}}}$; 2)~the efficiency due to propagating field absorption ${\eta_{{\rm{propagation}}}}$; and 3) the efficiency due to reflections at the body--air interface ${\eta_{{\rm{reflections}}}}$. Specifically, the total radiation efficiency ${\eta_{\rm{total}}}$ could be expressed as
\begin{equation}\label{eq:approx}
\begin{aligned}
{\eta_{\rm{total}}} &= {P_{\rm{air}}}/{P_{\rm{impl}}}  \\
&\approx{\eta_{\rm{near~field}}} \cdot {\eta_{\rm{propagation}}} \cdot {\eta_{\rm{reflections}}},
\end{aligned}
\end{equation}

\noindent where ${\eta_{\rm{total}}}$ represents the total radiation efficiency, $P_{\rm{air}}$ is the total power entered into the body tissue, and $P_{\rm{impl}}$ is the total power reaching free space. Unlike real-life antennas, the radiation efficiency of interest excludes the ohmic loss of the antenna itself.

The closed-form expressions for various efficiencies can be approximated using the operating frequency, the permittivity of body tissue, and the dimension of the body model \cite{skrivervik2019fundamental, Gao_2023}. In the case of TM sources, the approximation expressions of the three efficiencies are:
\begin{equation}\label{eq:elenear}
\begin{aligned}
&{\eta_{\rm near~field}} \\
& \mathop\approx \frac{{\mu _0}^{\frac{3}{2}}\rm{Re} \left( {\sqrt {\varepsilon_{\rm{body}} } } \right){\omega ^3}+ {\rm Im} \left( {{{\varepsilon_{\rm{body}}^{ - 1} }}} \right) {\emph d}^{ - 3}}{{\mu _0}^{\frac{3}{2}}{\rm Re} \left( {\sqrt {\varepsilon_{\rm{body}} } } \right){\omega ^3} + {\rm Im} \left( {{{\varepsilon_{\rm{body}}^{ - 1} }}} \right) {\emph r_{\rm{impl}}^{ - 3}}},
\end{aligned}
\end{equation}
\begin{equation}
\begin{aligned}
 {\eta_{\text{propagation}}} & = {e^{ - 2k''_{\rm{body}} \left( {d - {r_{{\mathop{\rm impl}\nolimits} }}} \right)}} \\
& = {e^{2\sqrt {{\mu _0}} {\mathop{\rm Im}\nolimits} \left( {\sqrt {\varepsilon _{\rm{body}} } } \right)\omega \left( {d - {r_{{\mathop{\rm impl}\nolimits} }}} \right)}},
\end{aligned}
\end{equation}
\begin{equation}
{\eta_{\rm reflections}} = \frac{{{\mathop{\rm Re}\nolimits} \left( {{{\left| T \right|}^2}/{Z_{\rm air}}} \right)}}{{{\mathop{\rm Re}\nolimits} \left( {1/{Z_{{\mathop{\rm body}\nolimits} }}} \right)}},
T = \frac{{2{Z_{{\mathop{\rm air}\nolimits} }}}}{{{Z_{{\mathop{\rm air}\nolimits} }} + {Z_{{\mathop{\rm body}\nolimits} }}}},
\end{equation}
\noindent where $T$ represents the transmission coefficient at the body--air interface, and $Z_{{\rm body}}$ and $Z_{{\rm air}}$ are the wave impedance in the body tissue and air at the body--air interface, respectively. 
For $r_{\rm body} > \lambda _{\rm m} / 2\pi$ (i.e., the wavelength in the medium with $\varepsilon_{{\rm m}}$), the corresponding wave impedance can be approximated as the intrinsic impedance of the medium
\begin{equation}
{Z_{\rm m}} \approx {\zeta_{\rm m}} = \sqrt {{\mu _0}/\varepsilon_{{\rm m}}}.
\end{equation}
For $r_{\rm body} \le \lambda _{\rm m}/ 2\pi$, the corresponding wave impedance needs to be evaluated as
\begin{equation}
\begin{aligned}
& {Z_{\rm m}} \approx Z_{{\rm m},n}^{\rm TM} = {\mathop{i}\nolimits} \zeta _{\rm m}  \frac{{\hat H_{n}^{\prime (2)}(k_{\rm m} {r_{\rm body}})}}{{\hat H_{n}^{(2)}(k_{\rm m}  {r_{\rm body}})}}\\
& \mathop = \limits^{n = 1} \frac{{\mu _0^{\frac{3}{2}}\sqrt {\varepsilon _{\rm m}} {\omega ^2}r_{{\text{body}}}^2 - i{\mu _0}\omega {r_{{\text{body}}}} - \sqrt {{\mu _0}/\varepsilon _{\rm m}} }}{{{\mu _0}\varepsilon _{\rm m}{\omega ^2}r_{{\text{body}}}^2 - i\sqrt {{\mu _0}\varepsilon_{\rm m} } \omega {r_{{\text{body}}}}}},
\end{aligned}
\end{equation}
\noindent where $Z_{{\rm m},n}^{\rm TM}$ denotes the spherical mode impedance for the TM source in the considered medium, and $\hat H_n^{(2)}$  is the Schelkunoff-type spherical Hankel function of the second kind and of order $n$ \cite{harrington1961time}. For most implants, $n = 1$ is considered to excite the dominant spherical mode.

In the case of TE sources, some inevitable changes are required in the approximation of near-field loss and reflection loss:
\begin{equation}\label{eq:magnear}
\begin{split} 
&{\eta_{\rm near~field}} \\
&\mathop\approx  \frac{{\mu _0}\left| { \varepsilon_{\rm body} } \right|\omega - 2\sqrt {{\mu _0}} \rm{Im} \left( {\sqrt { \varepsilon_{\rm body} } } \right){\emph d}^{ - 1}}{{\mu _0}\left| {\varepsilon_{\rm body} } \right|\omega  - 2\sqrt {{\mu _0}} \rm{Im} \left( {\sqrt { \varepsilon_{\rm body} } } \right){\emph r}_{impl}^{ - 1}},
\end{split}
\end{equation}
\begin{equation}
\begin{split} 
Z_{{\rm m},n}^{\rm TE} & = - {\mathop{i}\nolimits} \zeta_{\rm med} \frac{{\hat H_{n}^{(2)}(k_{\rm m} {r_{\rm body}})}}{{\hat H_{ n}^{\prime (2)}(k_{\rm m}  {r_{\rm body}})}} \\
& \mathop= \limits^{n = 1} \frac{{\mu _0^{\frac{3}{2}}\sqrt {\varepsilon_{\rm m} } {\omega ^2}r_{\rm body}^2 - i{\mu _0}\omega {r_{\rm body}}}}{{{\mu _0} \varepsilon_{\rm m} {\omega ^2}r_{\rm body}^2 - i\sqrt {{\mu _0} \varepsilon_{\rm m} } \omega {r_{\rm body}} - 1}},
\end{split}
\end{equation}
\noindent where $Z_{{\rm m},n}^{\rm TE}$ denotes the spherical mode impedance for the TE source in the considered medium. For $r_{\rm body} \le \lambda _{\rm m}/ 2\pi$, ${Z_{\rm m}} \approx Z_{{\rm m},n}^{\rm TE}$.

\subsection*{Closed-Form Approximation of Path Loss and Gain}
A more general model of implanted sources, especially for large host bodies, is an offset implanted source within a spherical body model, i.e., the source implantation depth $d$ is less than the curvature radius of the body-air interface $r_{\rm body}$. The considered body-air interface represents the one the closest to the implant, i.e. the distance from the implant to this interface is the implantation depth. For large host bodies such as the human abdomen, its curvature radius $r_{\rm body}$ is significantly greater than the implantation depth $d$.

For an offset implanted source within a homogeneous spherical body model, it is complicated to directly assess the total radiation efficiency because the propagating field loss varies in different directions. In such instances, the wireless link of greatest concern becomes the one in the direction from the implant to the closest body-air interface.

\begin{figure}[!b]% (b,c) -> el. mag sources + caption
\centering
\includegraphics[scale=0.54]{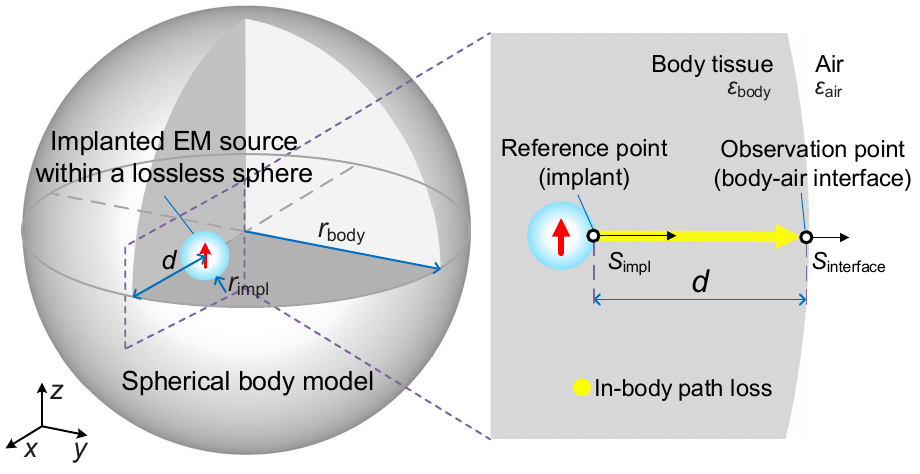}
\caption{\label{fig1} (a) View of the formulated spherical body model: a stratified spherical body model of dispersive body tissue with an implanted source surrounded by a small lossless sphere. For a typical body phantom, 1 is the muscle layer, 2 is the fat layer and 3 is the skin layer. The distribution of (b) electric fields \textbf{E} for a TM source and (c) magnetic fields \textbf{H} for a TE source.}
\end{figure} 

To characterize the link loss in the direction of interest, another efficiency indicator becomes critical, namely, the in-body path loss $\eta_{\rm in-body}$ \cite{Gao_2023}. This indicator quantifies the losses in power density of EM waves propagating in the body through the shortest wireless link, thus giving a maximal bound to assess the link efficiency between an implantable source and an on-body external receiver. Note that the considered spherical body phantom has an electrically large curvature radius of the body-air interface (i.e., $r_{\rm body}>\lambda _{\rm body}$), to avoid resonance effects that small body models may induce \cite{Gao_2024}.

The in-body path loss can be approximated using a closed-form expression similar to (\ref{eq:approx})
\begin{equation}\label{eq:approx_link}
\begin{aligned}
{\eta_{\rm{in-body}}} &= \frac{{{\rm Re} \left({S_{\rm{body}}} \right)}\cdot d^2} {{{\rm Re} \left({S_{\rm{impl}}} \right)}\cdot r_{\rm impl}^2} \\
&\approx{\eta_{\rm{near~field}}} \cdot {\eta_{\rm{propagation}}} \cdot {\eta_{\rm{reflections}}},
\end{aligned}
\end{equation}

\noindent where ${\rm Re}(S_{\rm{body}})$ and ${\rm Re}(S_{\rm{impl}})$ represent the radial component (i.e., normal to the body interface) of the radiated power density at the body interface and the implant encapsulation, respectively, $S$ represents the radial component of Poynting's vector $\mathbf{S}= \mathbf{E} \times \mathbf{H}^* $, and the factor ${d^2}/{r_{\rm impl}^2}$ accounts for the radial spreading of spherical EM waves. In the approximation of ${\eta_{\rm{in-body}}}$, ${\eta_{\rm{near~field}}}$ and ${\eta_{\rm{propagation}}}$ are consistent with those for radiation efficiency approximation; as for ${\eta_{\rm{reflections}}}$, ${Z_{\rm{air}}}$ and ${Z_{\rm{body}}}$ are directly taken as the intrinsic impedance in the corresponding medium.

Furthermore, it is possible to approximate the gain of an implanted source by extending the in-body path loss to the EM radiation reaching the far-field region in free space. To achieve this, another loss contribution, the efficiency due to refractions at the body-air interface ${\eta_{\rm{refractions}}}$, needs to be accounted for \cite{Gao_2024_FS}. As a consequence, the gain of the implanted source can be expressed as
\begin{equation}\label{eq:approx_gain}
\begin{split} 
&{G_{\rm{impl}}} \approx G_{\rm{intrinsic}} \cdot {\eta_{\rm{near~field}}} \cdot {\eta_{\rm{propagation}}} \cdot {\eta_{\rm{reflections}}} \\ 
&\cdot {\eta_{\rm{refractions}}},
\end{split} 
\end{equation}

\noindent where $G_{\rm{intrinsic}}$ represents the intrinsic gain of the implanted antenna, i.e., $G_{\rm{dip}} = {\rm 1.50}$ for an electrically short dipole. By analyzing the refraction of spherical waves, the closed-form approximation of ${\eta_{\rm{refractions}}}$ can be obtained as 
\begin{equation}
{\eta_{\rm{refractions}}} \approx \frac{{r_{{\text{body}}}^2}}{{{{\left[ {\frac{{{{k'}_{{\text{body}}}}}}{{{{k'}_{{\text{air}}}}}}\left( {{r_{{\text{body}}}} - d} \right) + d} \right]}^2}}}.
\end{equation}

\noindent In particular, for body models with a large curvature radius of the body-air interface (${r_{{\text{body}}}} \to \infty $), ${\eta_{\rm{refractions}}}$ can be further simplified as
\begin{equation}
{\eta_{{\text{refraction}}}} \approx \frac{{\varepsilon}_{0}}{{\varepsilon '}_{{\text{body}}}}.
\end{equation}

\subsection*{Link Budget of Implantable Wireless Systems}
In designing wireless systems, it is crucial to assess the link budget, which takes into account all of the power gains and losses from the transmitter to the receiver. This ensures stable and efficient wireless connectivity between an implantable RF device and an external unit. Similar to most wireless RF systems, the link budget can be effectively evaluated for prior design evaluation \cite{Agarwal_2017}.

For wireless systems where the implant acts as the transmitter, the received power at the external unit (or base station) is calculated and expressed logarithmically as \cite{Sani_2009}
\begin{equation}\label{eq:linkbudget1}
{P_{\text{ext} }} = {P_{{\text{impl}}}} + {G_{{\text{impl}}}} - PL + {G_{{\text{ext}}}},
\end{equation}
 where $P_{\text{ext} }$ is the received power at the external unit (receiver), $P_{{\text{impl}}}$ is the power radiated by the implant (transmitter), $G_{{\text{impl}}}$ and $G_{{\text{ext}}}$ are the total gains of the implant and the external unit, respectively, and $PL$ represents the free-space path loss. $G_{{\text{impl}}}$ takes into account the radiation of the implant under the influence of the host body, as the host body is part of the implanted antenna that radiates EM waves from the body-air interface. Because of the considerable absorption losses within the host body, $G_{{\text{impl}}}$ experiences substantial attenuation upon implantation compared to its achievable radiation in free space. Common applications of such cases are implants acting as active responders to communicate wirelessly with external devices\cite{Yoo_2021}.

Similarly, for wireless systems where the implant acts as the receiver, the received power at the implant is calculated and expressed logarithmically as
\begin{equation}\label{eq:linkbudget2}
{P_{\text{impl} }} = {P_{{\text{ext}}}} + {G_{{\text{ext}}}} - PL + {G_{{\text{impl}}}},
\end{equation}
where $P_{\text{impl} }$ is the received power at the implant, and $P_{{\text{ext}}}$ is the power radiated by the external unit (transmitter). $G_{{\text{impl}}}$ is also included in the calculation of this link budget. Typical applications of such cases could be RF wireless power transfer to implants or backscattering communications excited by external readers \cite{Yoo_2021}.

For both cases above, the power of the transmitter is limited by the FCC regulations \cite{federal1993understanding}, and the Specific Absorption Rate (SAR) assessment for wireless systems of implantable RF devices is required to meet the specification in IEEE C95.1 2019 standard \cite{IEEEStandard}.

\section*{Results}% sources
\subsection*{Mesoscopic Mechanisms of Radiation Efficiency}
\begin{figure}[!b]
\centering
\includegraphics[scale=0.495]{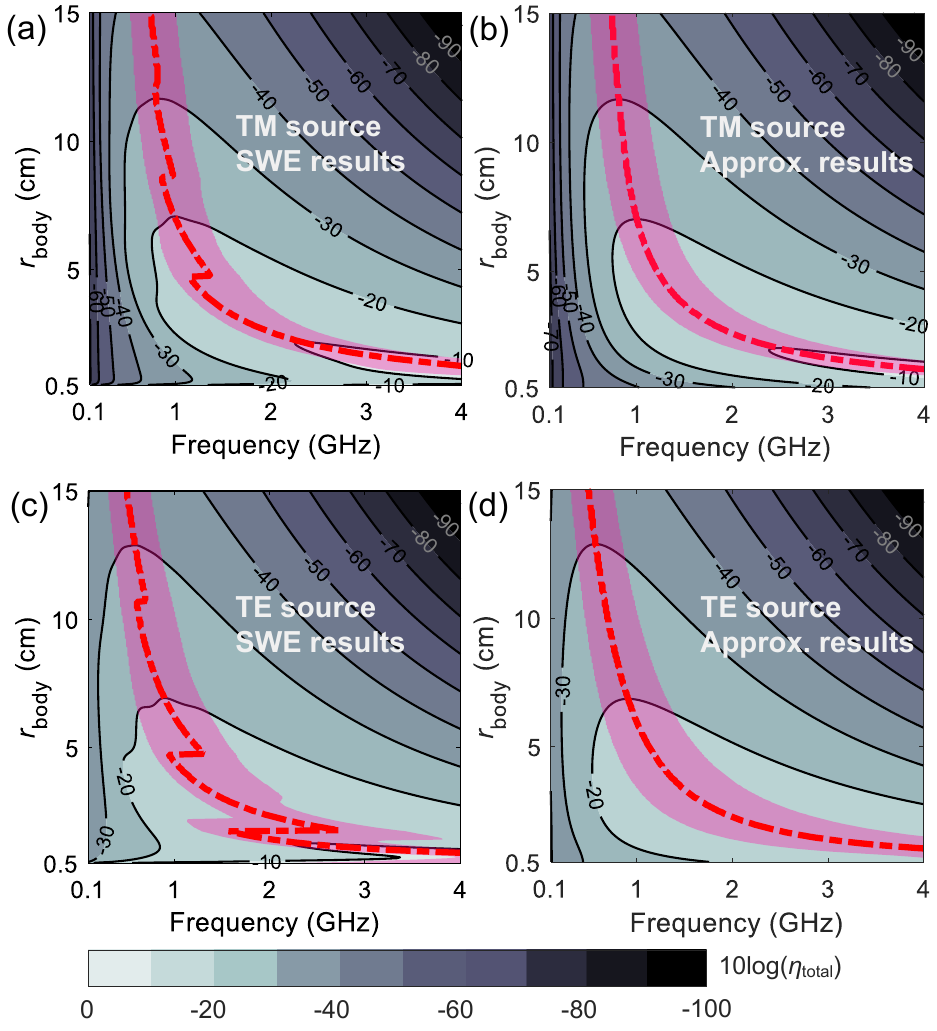}
\caption{\label{fig3} Distributions of $\eta_{\rm total}$ as a function of $f$ and $r_{\rm body}$. Different source types and calculation methods are applied, which are annotated in the subfigures. The curves of $f_{\rm opt}$ are drawn in red dot-dashed curves, and the surrounding shaded area shows the acceptable range of $f_{\rm opt}$ when the optimal tolerance of $\eta_{\rm total}$ is 0.8. Similar demonstrations are in the following figures if not indicated otherwise.}
\end{figure}

To improve the efficiency of implantable RF devices, a deep understanding of loss mechanisms is required. In this section, the physical mechanisms of the three loss contributions are investigated by looking into multiple key parameters, and the optimal frequency $f_{\rm opt}$ is introduced for the assessment and optimization of radiation efficiency $\eta_{\rm total}$. Note that the optimal frequency $f_{\rm opt}$ refers to the operating frequency that maximizes $\eta_{\rm total}$ of the implanted source. Using radiation efficiency as a metric allows for a comprehensive characterization of the implant's overall EM radiation, regardless of direction.

We start by considering a spherical body model with a centered source, where $r_{\rm body} \in [0.5, 15]$~cm and $r_{\rm impl} = 0.5$~cm if not otherwise specified. In this scenario, the implantation depth of the source $d = r_{\rm body}$. To simplify the studied model, a homogeneous spherical phantom made of muscle is studied to give conservative results of radiation efficiencies $\eta_{\rm total}$. 

For $f \in [0.1, 4]$~GHz, $\eta_{\rm total}$ of the implanted source is demonstrated as a function of $f$ and $r_{\rm body}$ in Fig.~\ref{fig3}. Whether for the TM or TE sources, $\eta_{\rm total}$ computed by the closed-form approximation method, as shown in Fig.~\ref{fig3} (b) and (d), are in good agreement with the results obtained with SWE method, as shown in Fig.~\ref{fig3} (a) and (c). Only when $r_{\rm body}$ is small or comparable to the wavelength in the body tissue $\lambda_{\rm{body}}$, the reflected waves (which are not attenuated completely) may turn the body phantom into a lossy dielectric resonator due to the presence of standing waves at certain frequencies \cite{Gao_2024}. This could explain the ripples observed in the curves of $\eta_{\rm total}$ in Fig.~\ref{fig2} when $r_{\rm body}=5$~cm. Due to this effect, there is a minor deviation of the output wave impedance at the body--air interface, which is not taken into account in the evaluation of ${\eta_{\rm reflections}}$. Overall, the closed-form approximation is effective in predicting the range of optimal frequencies for various implantation conditions. The results qualitatively indicate that $f_{\rm opt}$ is in the sub-GHz range or GHz-range for cm-sized spherical body phantom; $f_{\rm opt}$ is approximately inversely proportional to $r_{\rm body}$.

\begin{figure}[!t]
\centering
\includegraphics[scale=0.64]{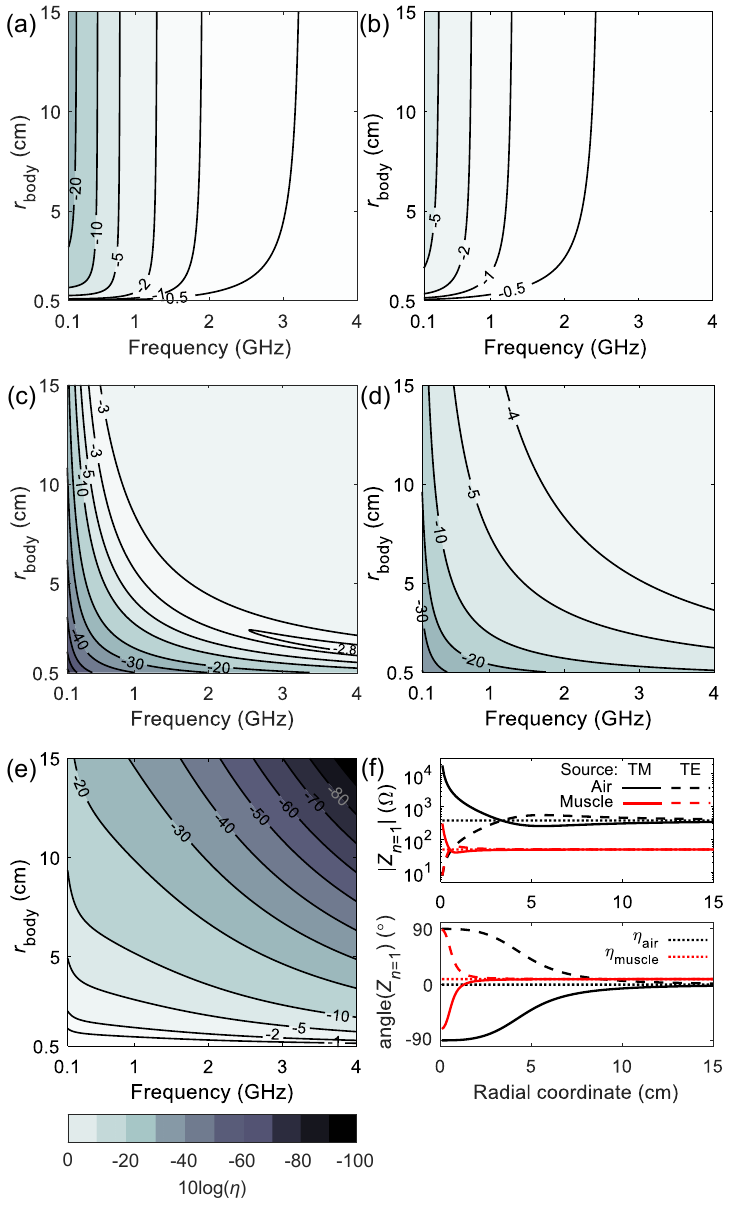}
\caption{\label{fig4} (a)--(e) Distributions of $\eta$ as a function of $f$ and $r_{\rm body}$ due to different loss contributions: $\eta_{\rm near~field}$ for (a) the TM and (b) TE source cases;  $\eta_{\rm reflections}$ for (c) the TM and (d) TE source cases; (e) $\eta_{\rm propagation}$ (same results for the TM and TE sources). (f) Magnitude and angle of $Z_{n=1}$ for different sources (TM and TE sources) and media (air and muscle) at $f$ = 1.2 GHz.} 
\end{figure}

Decomposing $\eta_{\rm total}$ into three main loss contributions enables an analytical understanding of the variation rules for different input parameters. As shown in Fig.~\ref{fig4} (a)--(e), losses due to different mechanisms exhibit distinctive characteristics as a function of $f$ and $r_{\rm body}$, where the source type also acts as a major factor for ${\eta_{\rm near~field}}$ and $\eta_{\rm reflections}$.

In the case of TM sources, the effect of $\eta_{\rm near~field}$ is prominent when $f < 1$~GHz, as shown in Fig.~\ref{fig4}(a), and its attenuation increases significantly as $f$ decreases (e.g., $\eta_{\rm near~field} \approx 0.009$  at $f = 0.2$~GHz and $\eta_{\rm near~field} \approx 0.104$ at $f = 0.5$~GHz when $r_{\rm body} = 15$~cm). For a relatively large $r_{\rm body}$, the approximate expression of $\eta_{\rm near~field}$ is no longer a function of $r_{\rm body}$ but a function of $r_{\rm impl}$. It is noted that both expressions of $\eta_{\rm propagation}$ and $\eta_{\rm reflections}$ can be approximated as a function of $f \cdot r_{\rm body}$ (or $\omega \cdot r_{\rm body}$) when the criterion $r_{{\rm{impl}}} \ll r_{\rm body}$ is met. The term $f \cdot r_{\rm body}$ is also proportional to the electrical length of $r_{\rm body}$, i.e., ${r_{\rm body}}/\lambda_{\rm{body}}  \approx f \cdot {r_{\rm body}}/\sqrt {{\mu _0}\varepsilon '}$. Due to the dominance of $f \cdot r_{\rm body}$, the contours in Fig.~\ref{fig4}(c)--(e) are similar to a group of curves with inverse proportionality (deviations are mainly caused by dispersive $\varepsilon _{\rm{body}}$). With the increase of $f \cdot r_{\rm body}$, $\eta_{\rm propagation}$ decays exponentially. In Fig.~\ref{fig4}(c), when $f \cdot r_{\rm body}$ is small enough (i.e., $r_{\rm body}$ is within the near-field region of the implant), attenuation due to $\eta_{\rm reflections}$ becomes significant as $Z_{{\mathop{\rm air}\nolimits},n = 1}^{{\mathop{\rm TM}\nolimits}}$ presents a high-impedance state, as illustrated in Fig. ~\ref{fig4}(f). With the increases of the electrical length of $r_{\rm body}$, $Z_{{\mathop{\rm air}\nolimits},n = 1}^{{\mathop{\rm TM}\nolimits}}$ drops rapidly and finally tends to the intrinsic impedance of the body tissue $\zeta$, thus $\eta_{\rm reflections}$ approaches a constant, i.e., $\eta_{\rm reflections}\approx 0.432$, which is evaluated using the intrinsic impedance.

In the case of TE sources, due to the weaker near-field coupling to body tissue, the effect of $\eta_{\rm near~field}$ presents less loss compared to the TM source case  (e.g., $\eta_{\rm near~field} \approx 0.223$ at $f = 0.2$~GHz and $\eta_{\rm near~field} \approx 0.485$ at $f = 0.5$~GHz when $r_{\rm body} = 15$~cm), as shown in Fig.~\ref{fig4}(b). Due to the difference between $Z_{{\rm air} ,n = 1}^{\rm TE}$ and $Z_{{\rm air} ,n = 1}^{\rm TM}$, as illustrated in Fig.~\ref{fig4}(f), for the same and small term $f \cdot r_{\rm body}$, the loss of $\eta_{\rm reflections}$ becomes moderate in Fig.~\ref{fig4}(d) compared with the results in Fig.~\ref{fig4}(c). As $f \cdot r_{\rm body}$ increases, the attenuation of $\eta_{\rm reflections}$ becomes greater than that of the TM source case, e.g., the area for 10log$(\eta_{\rm reflections}) < -5$ in Fig.~\ref{fig4}(d) is larger than that of Fig.~\ref{fig4}(c), until the efficiency slowly reaches a constant when $r_{\rm body}$ is in the far-field region.

\subsection*{Optimal Frequency: Trade-Offs between Implant Size, Scenarios, and Implantation Depth}\label{sec:optimal_frequency}
The optimal frequency $f_{\rm opt}$ is the best compromise between the three loss contributions. Among them, $\eta_{\rm propagation}$, i.e., the loss due to the propagating field, is inevitable and decreases as the electrical length of $d$ increases. However, due to the different mechanisms of near-field coupling and wave reflections, $\eta_{\rm near~field}$ and $\eta_{\rm reflections}$ only cause significant attenuation for certain implantation conditions of each. In this subsection, we first analyze the combinations of two selected loss contributions, i.e., combination C1: $\eta_{\rm near~field} \cdot \eta_{\rm propagation}$, and combination C2: $\eta_{\rm propagation} \cdot \eta_{\rm reflections}$. We then investigate the decision rules for $f_{\rm opt}$, which leads to a quick estimation method to assess the optimal choices of operating frequency.

\begin{figure}[!t]
\centering
\includegraphics[scale=0.495]{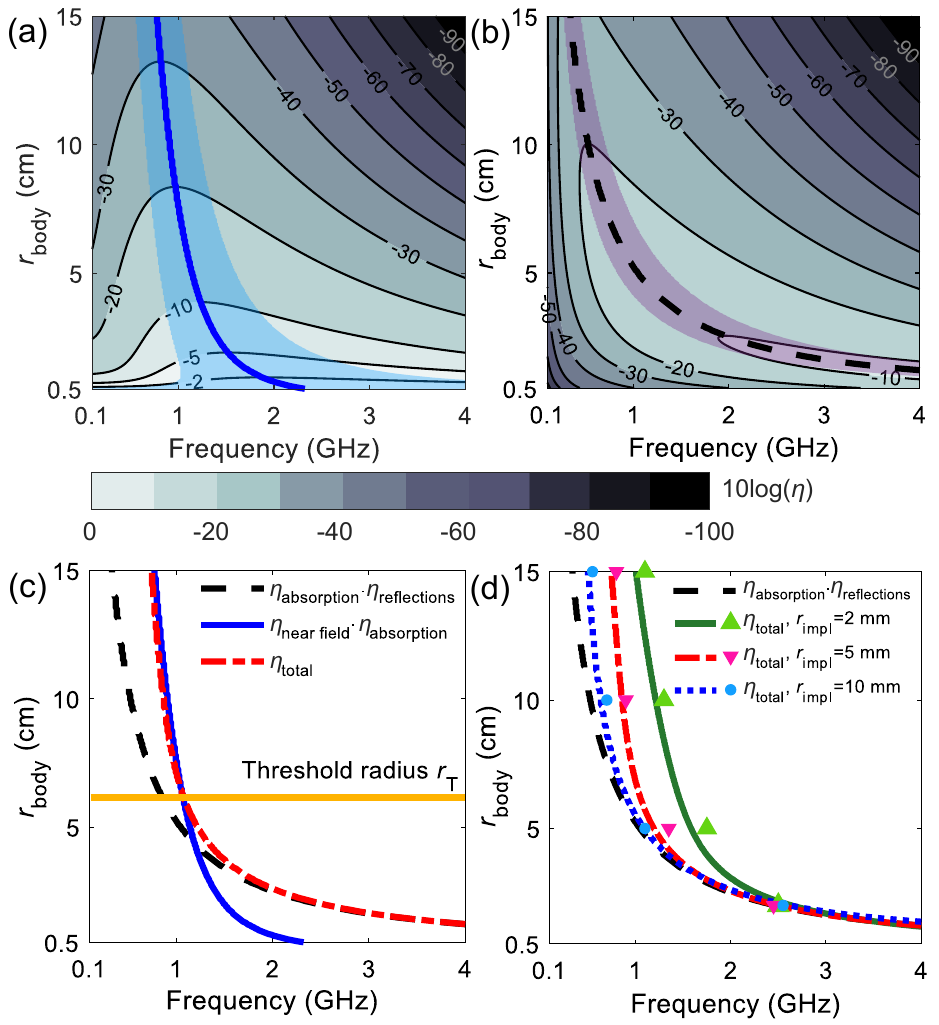}
\caption{\label{fig5} (a),(b) Distributions of $\eta$ due to (a) C1 and (b) C2 as a function of $f$ and $r_{\rm body}$ for the TM source case. (c) Curves of $f_{\rm opt}$, $f_{\rm opt,C1}$ and $f_{\rm opt,C2}$ for the TM source case, in which $r_{\rm T}$ is introduced. (d) Comparison of $f_{\rm opt}$ for TM sources with different $r_{\rm impl}$.}
\end{figure}

Taking the spherical body model with a centered TM source as the first scenario, Fig.~\ref{fig5}(a) and (b) demonstrate the efficiency results given by the combinations C1 and C2 with the corresponding curves of $f_{\rm opt, C1}$ and $f_{\rm opt, C2}$, respectively. Once these curves are plotted together with the curve of $f_{\rm opt}$ [results for the total efficiency, first shown in Fig.~\ref{fig3}(b)], as shown in Fig.~\ref{fig5}(c), we observe that the curves for optimal frequency are partially overlapped for different conditions. Thus, we intuitively define the $r_{\rm body}$ at the intersection point between curves of $f_{\rm opt}$ and $f_{\rm opt, C1}$ as the threshold curvature radius $r_{\rm T}$. The rules are then straightforward: when $r_{\rm body} > r_{\rm T}$, the curve of $f_{\rm opt}$ basically overlaps the curve of $f_{\rm opt, C1}$, i.e., C1 dominants the value of $f_{\rm opt}$ in this condition; on the contrary, when $r_{\rm body} \le r_{\rm T}$, C2 starts to dominate the value of $f_{\rm opt}$. Furthermore, since the reduction of $r_{\rm impl}$ leads to greater attenuation of $\eta_{\rm near~field}$, the overlapping portion between the curves of $f_{\rm opt}$ and $f_{\rm opt,C2}$ becomes smaller in Fig.~\ref{fig5}(d), i.e., $r_{\rm T}$ decreases. The above analytical results are validated numerically by a realistic source, denoted as additional points in Fig.~\ref{fig5}(d). See the Appendix for detailed descriptions.

\begin{figure}[t]
\centering
\includegraphics[scale=0.51]{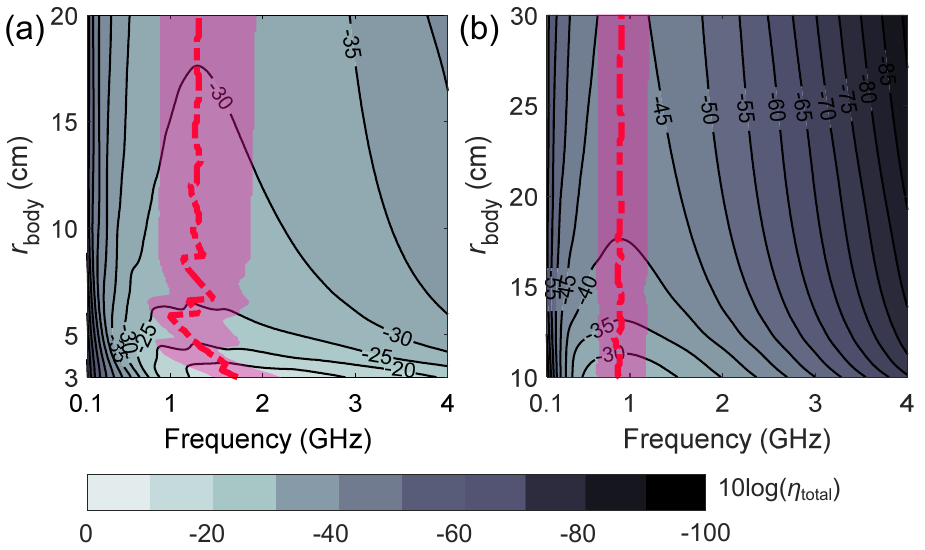}
\caption{\label{fig6}Distribution of $\eta_{\rm total}$ as a function of the frequency and $r_{\rm body}$ for the TM source case, where (a) $d=3$~cm and (2) $d=10$~cm.}
\end{figure}

More generally, for a spherical body model with an offset implanted source, further investigations reveal the variation of the optimal frequency $f_{\rm opt}$ with the curvature radius of the nearest body--air interface $r_{\rm body}$. As shown in Fig.~\ref{fig6}(a), by setting the implantation depth $d=r_{\rm body}-r_{\rm feed}=3$~cm, $\eta_{\rm total}$ is calculated by SWE method with the increase of $r_{\rm body}$. When $r_{\rm body} < r_{\rm T} \approx 6.15$~cm, $f_{\rm opt}$ starts to fluctuate and gradually decreases. In this scenario, a safe estimate of $f_{\rm opt}$ is to account for the three loss contributions and note the use of $d$ and $r_{\rm body}$ in the closed-form expressions (since the source is offset). As $r_{\rm body}$ increases further, $f_{\rm opt}$ tends to stabilize instead, and eventually $f_{\rm opt}$ becomes a constant at 1.31~GHz. This $f_{\rm opt}$ is lower than that of the centered source model. In terms of the loss mechanisms, the expression of $\eta_{\rm near~field}$ is still valid according to the equivalence principle, while the value of $\eta_{\rm reflections}$ becomes stable and could be approximated by using the intrinsic impedance directly. As the effect of $\eta_{\rm reflections}$ becomes weaker and tends to a constant, $f_{\rm opt}$ decreases and tends to $f_{\rm opt, C1}$.

It is worth noting that the increase in $r_{\rm body}$ could introduce more losses in $\eta_{\rm propagation}$ due to the increased body model, but its effect on $f_{\rm opt}$ is negligible as long as the implantation depth (i.e., the shortest path from the source to the interface) is fixed. According to this characteristic, $f_{\rm opt}$ can in fact be quickly assessed by the approximation of the in-body path loss. In particular, once the effect of $\eta_{\rm reflections}$ is negligible for large $r_{\rm body}$, the loss combination C1, i.e., $f_{\rm opt,C1}$ in Fig.~\ref{fig5}(a), becomes effective to determine $f_{\rm opt}$. Following the case of $d=3$~cm, the corresponding $f_{\rm opt,C1}$ in Fig.~\ref{fig5}(a) at $r_{\rm body}=3$~cm is 1.34~GHz, which is close to the stabilized $f_{\rm opt}$ in Fig.~\ref{fig6}(a). The same rules can also be applied in scenarios with larger implantation depths. As shown in Fig.~\ref{fig6}(b), for $d=10$~cm, since this implantation depth is greater than $r_{\rm T}$, $f_{\rm opt}$ remains almost the same with the increase of $r_{\rm body}$. Thus, $f_{\rm opt}$ is approximately equal to $f_{\rm opt, C1}$, which provides a quick estimation approach. Once $r_{\rm body}$ is large enough, the spherical body model becomes close to a planar body phantom in which the source is implanted at the same depth. A similar topic was investigated in depth in \cite{poon_optimal_2010}, which delves into wireless power transfer into biological tissues.

\begin{figure}[!t]
\centering
\includegraphics[scale=0.5]{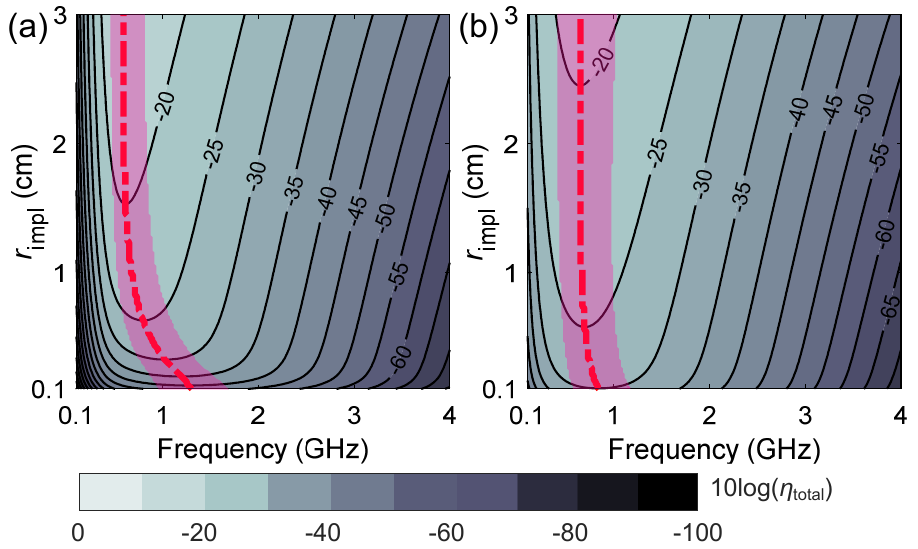}
\caption{\label{fig7} Distributions of $\eta_{\rm total}$ as a function of $f$ and $r_{\rm impl}$ for (a) the TM and (b) TE sources via the approximation method, where $r_{\rm body}=10$~cm. }
\end{figure}

The effect of $r_{\rm impl}$ on $f_{\rm opt}$ is demonstrated in Fig. \ref{fig7}, where $\eta_{\rm total}$ is evaluated as a function of $f$ and $r_{\rm impl}$ for a centered source within a $r_{\rm body}=10$~cm spherical body model. As indicated in Eq.~(\ref{eq:elenear}) and (\ref{eq:magnear}), regardless of the source type, the larger the source is, the higher $\eta_{\rm total}$. For a fixed implantation depth $d=r_{\rm body}=10$cm, the threshold value for $r_{\rm impl}$ is worth noting. In the case of TM sources [Fig.~\ref{fig7}(a)], when the implant size $r_{\rm impl}$ is smaller than around 1 cm, the aggravating near-field loss leads to the increase of $f_{\rm opt}$. Once $r_{\rm impl}$ is large enough, $f_{\rm opt}$ tends to a constant value as the near-field loss is no longer dominant (i.e., C2 dominates with a nearly constant $f \cdot r_{\rm body}$). In the case of TE sources [Fig.~\ref{fig7}(b)], the threshold value of ${r_{\rm impl, T}}$ changes to around 0.5~cm due to the weaker near-field coupling of the TE source, above which $f_{\rm opt}$ can be seen as a constant.

In summary, $f_{\rm opt}$ is determined by the trade-off among the three main loss contributions. As the operating frequency increases, $\eta_{\rm propagation}$ becomes dominant due to the increase in the electrical length of $d$ and the dispersion property of body tissues. Conversely, as the operating frequency decreases, $\eta_{\rm near~field}$ and $\eta_{\rm reflections}$ are likely to dominate. They are functions governed by the implant encapsulation size and the curvature radius of the body-air interface, respectively. In the initial stage of implantable antennas, an affordable and wise strategy is to first estimate the optimal frequency and then use the closest licensed frequency band. This can help effectively improve the implant's radiation performance and thus increase the wireless link budget.

Practically, for most implants in the human body (i.e., $r_{\rm body}$ is electrically large), $f_{\rm opt}$ is dominated by $\eta_{\rm near field}$ and $\eta_{\rm propagation}$, i.e., $f_{\rm opt}=f_{\rm opt, C1}$. This means that the most efficient way to determine the optimal frequency under this condition is to optimize the unavoidable in-body path loss. A numerical example is given in the subsection "Numerical Demonstration".

\subsection*{Maximum Achievable EM Efficiency and~Source Optimization}
The maximum achievable efficiency is one of the key indicators for the design of wireless implantable bioelectronics. To develop the design strategies for maximizing wireless efficiency, quantifying trade-offs between different loss contributions is necessary. 

\begin{figure}[!t]
\centering
\includegraphics[scale=0.59]{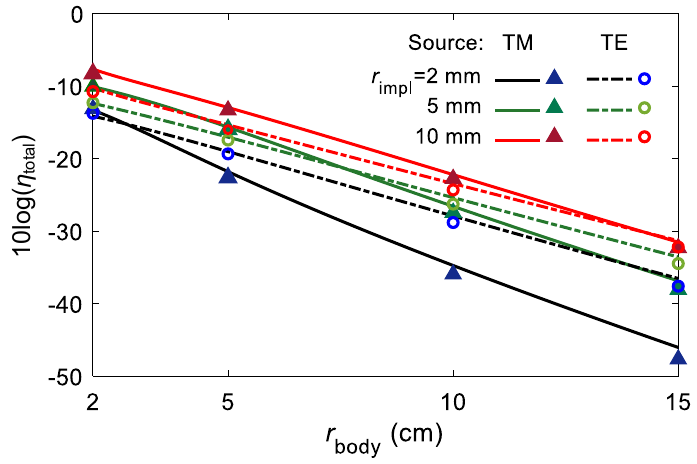}
\caption{\label{fig8} Maximum achievable $\eta_{\rm total}$ exponentially decays with $r_{\rm body}$, where both  TM and TE sources are analyzed for different $r_{\rm impl}$.}
\end{figure}

To investigate the effects of different host body dimensions, we first considered a spherical body model with a centered source. In this scenario, the maximum achievable efficiency is obtained as the maximum $\eta_{\rm total}$ at $f_{\rm opt}$, which is a function of the body radius $r_{\rm body}$ (i.e., the implantation depth $d$) and the implant encapsulation size $r_{\rm impl}$. As shown in Fig.~\ref{fig8}, the maximum achievable $\eta_{\rm total}$ is computed for both TM and TE sources. The optimal choice of the source type is related to both the implant encapsulation size $r_{\rm impl}$ and $r_{\rm body}$: the larger $r_{\rm body}$, the smaller $f_{\rm opt}$ is, which makes $\eta_{\rm near~field}$ smaller and further promotes the attenuation of $\eta_{\rm total}$, especially for the TM source case or the source with a small $r_{\rm impl}$. On the other hand, for a body model with a small $r_{\rm body}$, the difference in $\eta_{\rm reflections}$ (illustrated in Fig.~\ref{fig4}(c) and (d)) makes the TE source slightly inferior to the TM source once the implant has a larger size ($r_{\rm impl}$). As a result, for scenarios of implants within small host bodies, (i.e., the curvature radius of the body-air interface $r_{\rm body} \le \lambda _{\rm body}$), the selection of electric or magnetic sources requires a specific analysis of the implantation location (depth, body dimensions, and tissue properties) and encapsulation size, as there is no absolute superiority among them. The evaluated analytical results are again verified effectively with the numerical results of the realistic source model (see Appendix), shown as discrete points in Fig.~\ref{fig8}.

\begin{figure}[!t]
\centering
\includegraphics[scale=0.61]{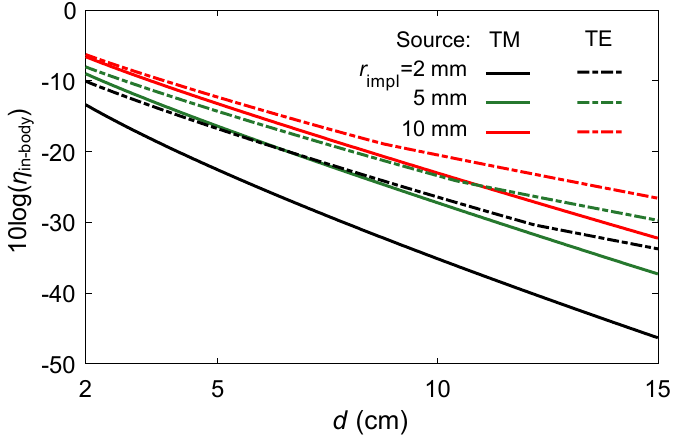}
\caption{\label{fig9} Maximum achievable $\eta_{\rm{in-body}}$ exponentially decays with the implantation depth $d$, where both  TM and TE sources are analyzed for different $r_{\rm impl}$.}
\end{figure}

However, for scenarios of implants within large host bodies (i.e., $r_{\rm body} > \lambda _{\rm body}$), an offset implanted source within a large spherical body model is considered. Here, the maximum achievable efficiency is obtained as the in-body path loss $\eta_{\rm in-body}$, which can be again calculated using closed-form expressions. Compared to the centered source scenario, here the intrinsic impedance is used to calculate $\eta_{\rm reflections}$, thus affecting the assessment of $f_{\rm opt}$ and the corresponding efficiency. Taking similar parameters in Fig.~\ref{fig8} but replacing $r_{\rm body}$ with the implantation depth $d$, the maximum achievable $\eta_{\rm{in-body}}$ are demonstrated in Fig.~\ref{fig9} as a function of $d$.

As a consequence, the TE sources (i.e., magnetic type sources) consistently achieve higher link efficiency due to the nature of weaker near-field coupling with body tissue compared to the electric sources. Quantitatively, their differences in $\eta_{\rm in-body}$ become significant for smaller encapsulation sizes and deeper implantation depths. In practical applications, TM sources (i.e., electric type sources) are suitable for shallow implants (such as subcutaneous implants) owing to the design flexibility (e.g., easy impedance matching and various shapes). As for magnetic type sources, they are suitable for deep implants, especially advantageous in the limitation of miniaturized encapsulation size. Similar conclusions can be drawn if the gain is used as a measure for offset implanted source scenarios. The only added refraction loss ${\eta_{\rm{refractions}}}$ has a limited influence on $f_{\rm opt}$, as it is affected by the dispersion properties of body tissues.

\section*{Discussion}
In this section, we discuss and demonstrate the effectiveness and application of the proposed design strategies through both numerical and experimental studies.

\subsection*{Numerical Demonstration}
\begin{figure}[t]
\centering
\includegraphics[scale=0.46]{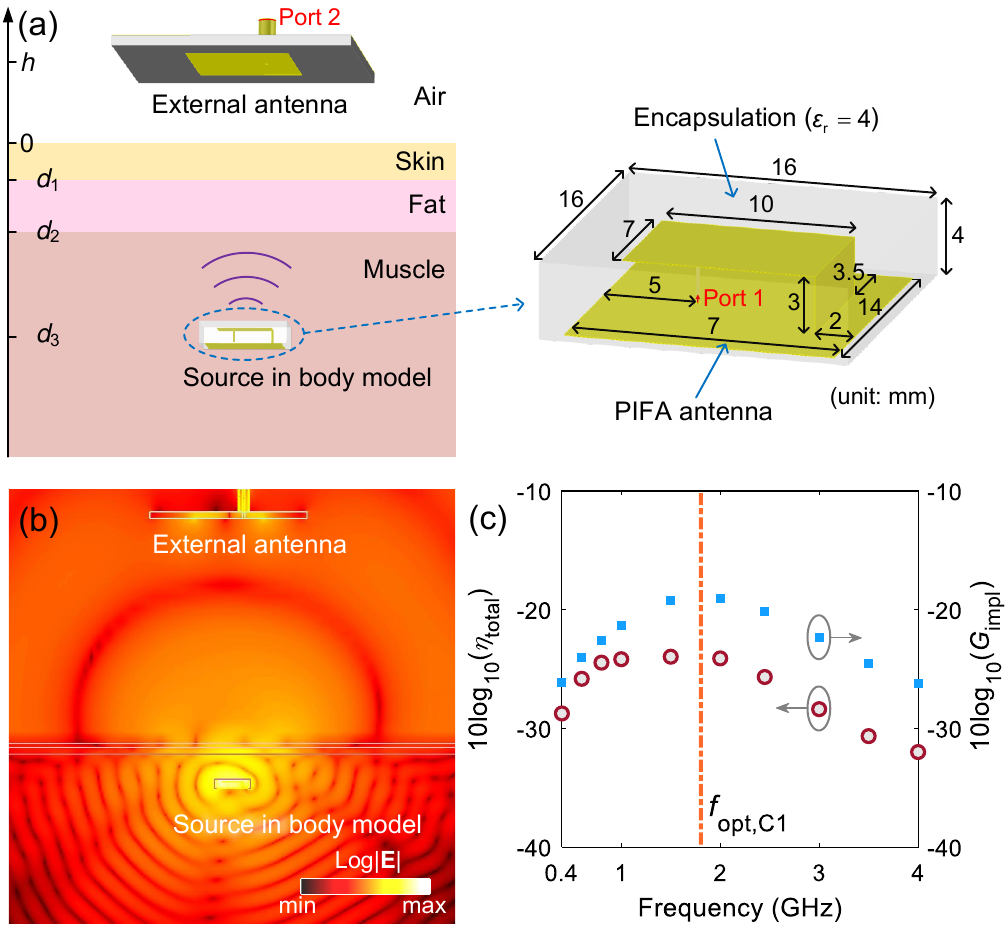}
\caption{\label{fig10}(a)~Schematic illustration of a wireless system for an implantable cardiac device. The implant in the body model is the transmitter while the external antenna is the receiver. (b)~Electric-field distribution at 2.45~GHz illustrating the EM radiation from the implant and wave propagation in different media. (c)~Simulated radiation efficiency and gain as a function of operating frequency, where the frequencies of optimal performance match well with the calculated $f_{\rm opt, C1}$.}
\end{figure}

To numerically demonstrate the design and optimization of wireless implants, we analyze a numerical example shown in Fig.~\ref{fig10}(a). A wireless system for an implantable cardiac device is studied using a stratified planar body model. 

Specifically, the implant, as the transmitter, is a planar inverted-F antenna encapsulated in a cuboid medium (lossless dielectric with $\varepsilon_{\rm{r}} = 4$) representing a simplified model of a cardiac implant. To build a complete wireless link, the external antenna, as a receiver, is a patch antenna placed externally and directly above the implant with the height $h$ = 100 mm from the body-air interface. The body model is an air-skin-fat-muscle stratified planar body model with the layer thicknesses of $d_{\rm{1}}$ = 2 mm and $d_{\rm{2}}$ = 5 mm. The implant is located in the muscle layer with a total depth of $d_{\rm{3}}$ = 20 mm. 

We use CST Microwave Studio to conduct full-wave EM field simulations for individual antennas and the wireless system. Both Port~1 of the implant and Port~2 of the external antenna are first designed to 50 $\Omega$. In analyzing the system at different operating frequencies, conjugate matching ports are applied, and the mismatch loss is excluded using the reflection coefficient. The conductor parts adopt perfect electric conductors to eliminate metal loss. The boundaries on the sides and bottom of the body model are set to open boundaries to reduce reflections from other body interfaces. An air layer with an open (add space) boundary is included above the body model to contain the external unit and for far-field radiation analysis. As illustrated in Fig.~\ref{fig10}(b), the EM waves radiated by the implant propagate as hemispherical waves in the air, which are partially received by the external antenna.

This numerical example can validate the quick estimation method of the optimal operating frequency $f_{\rm opt}$. As shown in Fig.~\ref{fig10}(c), $\eta_{\rm total}$ and $G_{\rm impl}$ of the implantable cardiac device are simulated for $f \in [0.4, 4]$~GHz. It is observed that both the maximum $\eta_{\rm total}$ and $G_{\rm impl}$ occur in the frequency range for $f \in [1.5, 2]$~GHz, i.e., $f_{\rm opt}(\eta_{\rm total}) \approx f_{\rm opt}(G_{\rm impl})$, which is consistent with the theoretical findings. Based on the quick estimation method for $f_{\rm opt}$, in this example, the implant under study can be represented by a homogeneous body model made of muscle with $d$ = 15 mm and $r_{\rm impl}\approx$ 4 mm (the ground plane of the antenna doubles the effective height of the encapsulation). As illustrated in Fig.~\ref{fig10}(c), the the optimal operating frequency is estimated as $f_{\rm opt}=f_{\rm opt, C1}$= 1.79~GHz. Thus, the fast estimated value $f_{\rm opt}$ is in good agreement with the simulation results, validating the proposed closed-form expressions and optimization strategy.

This numerical example can be further employed to demonstrate the usability of the proposed closed-form expressions for assessing the gain and link budget of implantable antennas. Since the transmit power of the implant is not determined, the wireless link efficiency is analyzed here in place of the link budget, by taking the ratio of the received power $P_{\rm ext}$ at Port~2 to the transmitted power $P_{\rm impl}$ at Port~1. It can also be approximated as indicated in Eq.~(\ref{eq:approx_gain}) and Eq.~(\ref{eq:linkbudget1}). In this example, both the implantable antenna and the external antenna operate at 2.45 GHz, i.e., in the 2.4-GHz Industrial, Scientific, and Medical (ISM) band. As shown in Table~\ref{tab:table1}, the gain of the implantable antenna $G_{\rm impl}$ and the link efficiency of the wireless system $P_{\rm ext}/P_{\rm impl}$ are simulated and approximated, respectively. Again, the approximation of the antenna gain considers a homogeneous body model of the same dimensions as the previous analysis. Compared with the simulated $G_{\rm impl}$, the approximate $G_{\rm impl}$ is 3.47 dB lower, which is mainly attributed to the estimation of the implant size in the calculation and the reduction of reflection loss caused by the fat-skin layer due to the matching layer effect. Using the simulated gain of the external antenna, the link efficiency of the entire wireless system is also approximated, which is only 1.78 dB lower than the simulation result. In summary, the above results confirm the efficiency of the proposed design method of implantable antennas achieved through an analytical approach.

\begin{table}[t]
  \begin{center}
    \caption{Comparison of simulated and approximate results for an implantable cardiac device.}
    \label{tab:table1}
    \begin{tabular}{l c c c c}
      \toprule % <-- Toprule here
      \hline
      \textbf{ } & \textbf{10log$(G_{\rm impl})$} & \textbf{10log$(G_{\rm ext})$} & $PL$ &\textbf{10log($\frac{P_{\rm ext}}{P_{\rm impl}}$)} \\ \hline % <-- Midrule here
      Simul. & --20.77 & 7.00 & -- & --35.68 \\
      Approx. & --24.24 & 7.00 & 20.22 & --37.46 \\ \hline
      \bottomrule % <-- Bottomrule here
    \end{tabular}
  \end{center}
\end{table}

\subsection*{Experimental Demonstration}
\begin{figure}[t]
\centering
\includegraphics[scale=0.48]{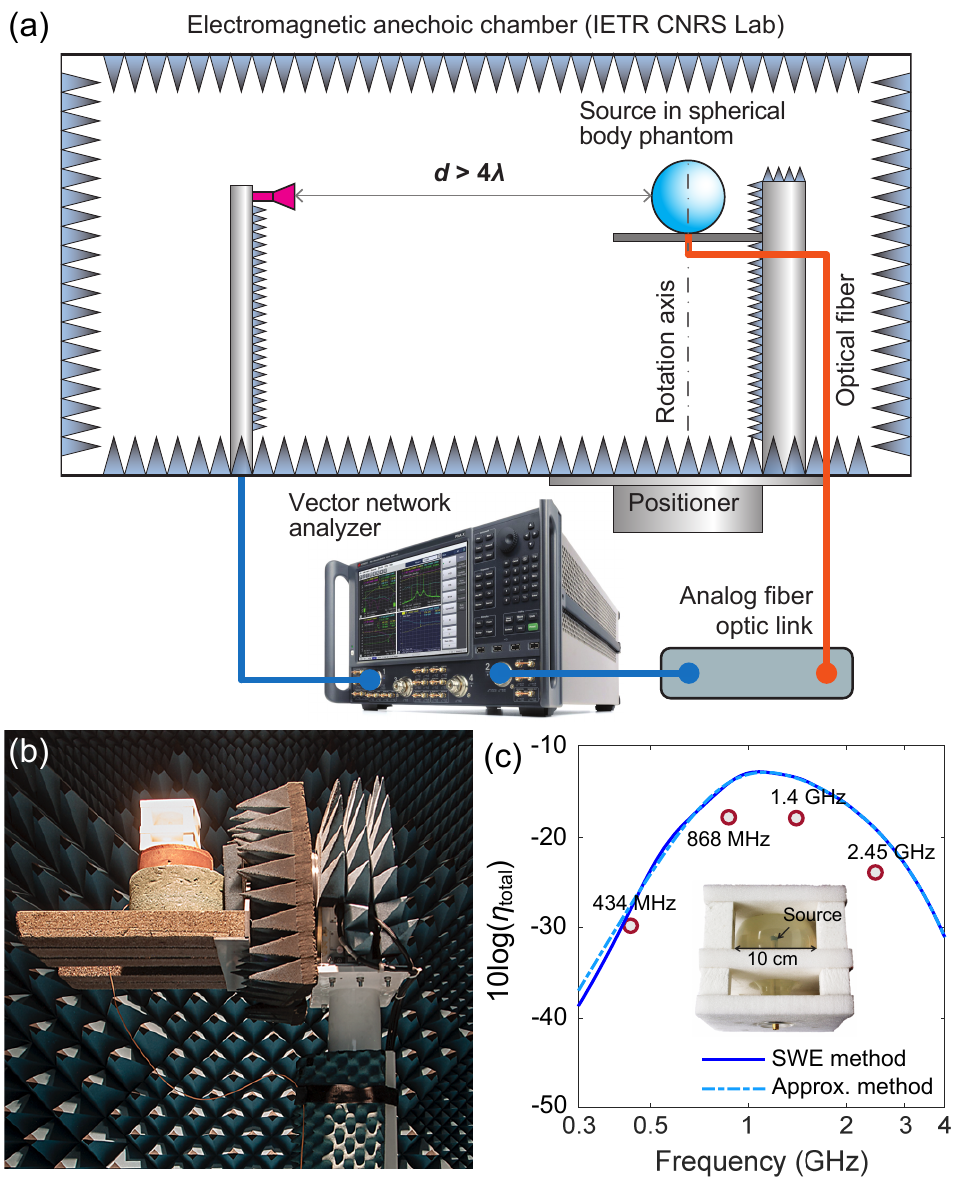}
\caption{\label{fig11}(a)~Outline of the experimental setup using an analog fiber-optic link. (b)~Source under test in a tissue-simulating model (phantom) mounted in the cm-wave anechoic chamber. (c)~Measured $\eta$ of the sources (designed based on the optimal criteria) compared to computed $\eta$ from the SWE and closed-form expressions. \textit{Inset}: spherical container with the implantable source and tissue-simulating liquid.}
\end{figure}

For the experimental demonstration of the radiation efficiency of implantable bioelectronics, we use the setup shown in Fig.~\ref{fig11}(a). In this $in~ vitro$ experiment, four capsule-conformal implantable antennas of the encapsulation size $19 \times \diameter 7$~mm were designed based on the optimal criteria and spanning the optimal frequency range. Specifically, the authorized frequency bands centered at 434, 868, 1400, and 2450 MHz were chosen for the demonstration. The achieved radiation efficiencies $\eta$ show about fivefold improvement compared to existing implantable devices of similar dimensions \cite{nikolayevAntennasIngestibleCapsule2016}. 

Measurement of small antennas in lossy liquids is a nontrivial task, which requires the use of non-standard characterization techniques to mitigate spurious radiation from the antenna feed. To address this challenge, we used a far-field direct illumination of the source in a spherical body phantom placed in a fully-anechoic chamber [Fig.~\ref{fig11}(b)] and fed with an electro-optical converter (enprobe LFA-3). The source is placed at the distance $d >4\lambda_{\rm{air}}$ (at 434 MHz) from the measurement horn fulfilling the far-field criterion for all frequencies of interest. The measured $\eta$ is estimated by dividing the measured antenna gain by the antenna directivity and correcting for the antenna reflection coefficient. Detailed descriptions of the antenna designs and the measurement procedure (including preparation of the tissue-simulating liquids) are presented in~\cite{nikolayevDielectricloadedConformalMicrostrip2019, nikolayevReconfigurableDualbandCapsuleconformal2022}.

The analysis is performed for the case of a $\diameter 100$-mm body phantom, which approximates a 5-cm implantation depth. Fig.~\ref{fig11}(c) compares the measured $\eta$ of the realistic sources with the computed $\eta$ from the SWE and closed-form expressions. The measured results are in close agreement with the theoretical ones and follow a similar trend. The efficiency peaks for 868-MHz and 1.4-GHz sources are close to the optimal frequency ranges.

The gap between the maximum achievable efficiency and the experimentally obtained ones is attributed mainly to 1) the material losses (note that the mismatch loss was de-embedded from the measurements), 2) the available surface area of the implant to radiate EM waves, and 3) presence of higher order spherical harmonics that increase losses. The 434-MHz design spans the entire available space on the cylindrical substrate and radiates in electric dipole mode, therefore it is the closest to the fundamental limits. Higher-frequency sources have smaller apertures and, most importantly, produce higher-order modes that increase the near-field loss. Despite these larger losses, obtained efficiencies of the 868-MHz and 1.4-GHz sources are higher than the 434-MHz source for the considered implantation with $d = 5$~cm. These results illustrate how the obtained design strategies can predict the optimal frequencies and source types for a given implantation depth and implant size. Finally, narrow higher-frequency sources operating at an optimal frequency range can be arranged in a circular array to increase the aperture size and to realize the adaptive beam-steering. Such a source will further minimize the energy dissipation in tissues, therefore maximizing its radiation efficiency.

\section*{Conclusions}

Wireless implantable bioelectronics have been rapidly growing to become an important set of tools in monitoring, assistance, and therapeutic treatment. This work investigates the electromagnetic radiation mechanism of wireless implantable devices and provides benchmarks to assess and optimize the radiation performance through an analytical approach. The overarching objective is to improve the EM efficiency of the implant, which is the dominant factor constraining the wireless link budget, thereby proposing design strategies and key rules for the preliminary design of implantable antennas.

Through in-depth analytical modeling of body-implanted electromagnetic sources, the EM efficiency is found to depend on multiple morphological and electromagnetic properties of the implant, host body, and operating frequency. Consequently, for a specific implantation site, achieving optimal radiation is possible through careful selection of the operating frequency and source type (electric or magnetic), in addition to adjusting the effective encapsulation size.

The optimal operating frequency is the best trade-off between the three loss contributions. As induced by the analytical model, the loss of the propagating field increases at higher frequencies, while near-field coupling and/or reflections at the body-air interface cause more severe losses at lower frequencies. By analyzing the characteristics of different loss contributions and their combinations, we propose quick estimation methods for the optimal frequency applicable to different implant scenarios. On the other hand, to maximize the radiation efficiency or link efficiency, we also analyze the effects of source type and encapsulation size and come up with design strategies for different implantation depths and body-air interface dimensions. 

The obtained design methods and strategies are further discussed in several numerical and experimental demonstrations of realistic implants, and the well-fitted estimation results reveal their value in practical applications. The proposed design rules and procedures are suitable in the early design stage of wireless implants to improve the link efficiency of the wireless system at the lowest possible cost. Further studies are required to more accurately assess the gain and link efficiency of implants with specific encapsulation geometries. 

\section*{Appendix}
\begin{figure}[!t]
\centering
\includegraphics[scale=0.485]{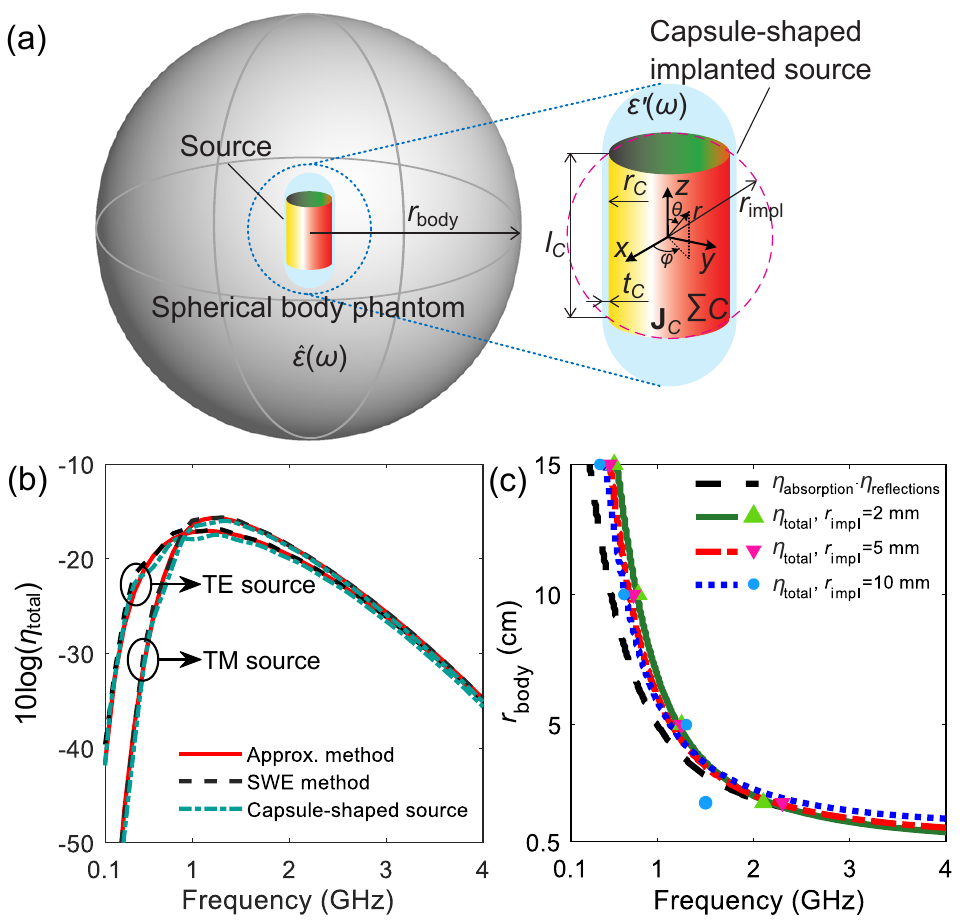}
\caption{\label{fig12}(a) View of the spherical model with a realistic capsule-shaped implanted source. (b) Frequency dependency of $\eta_{\rm total}$ for a TM or TE source with $r_{\rm impl} = 5$ mm implanted at the center point of a spherical body phantom with $r_{\rm body}=5$ cm. (c) Comparison of the curves of $f_{\rm opt}$ for TE sources with different $r_{\rm impl}$. }
\end{figure}

To validate the obtained theoretical results, we developed a realistic capsule-shaped model of an implantable source. As shown in Fig.~\ref{fig12}(a), the source is analyzed at the center of the spherical tissue model. Within the spherical phantom of complex permittivity $\varepsilon (\omega )$, the cylindrical surface of the capsule-shaped source $\Sigma_C$ is defined by the variable length $l_{\rm C}$ and radius $r_{\rm C}$. To represent a generic pill-shaped in-body device, a lossless region with the permittivity  $\varepsilon '(\omega )$ encloses the surface $\Sigma_C$. The region consists of a cylinder of length $l_{\rm C}$ and radius $r_{\rm C}+t_{\rm C}$ and two hemispheres of the same radius. To meet a similar source size in the canonical model of Fig.~\ref{fig1}, the circumradius of the source satisfies $r_{\rm impl} = \sqrt{l_{\rm C}^2/4 + (r_{\rm C}+t_{\rm C})^2}$. Consequently, 
if $t_{\rm C} \ll r_{\rm impl}$, the dimensions of $\Sigma_C$ are derived from $r_{\rm impl} \approx \sqrt{l_{\rm C}^2/4 + r_{\rm C}^2}$ as $l_{\rm C} = 6r_{\rm impl}/\sqrt {13}$, $r_{\rm C} = l_{\rm C}/3$, and $t_{\rm C}= r_{\rm impl}/10$.

The surface current density $\mathbf{J}_C=(J_{C,r}, J_{C,\varphi}, J_{C,z})$ on the cylindrical surface $\Sigma_C$ is defined for TM and TE sources as $\mathbf{J}_{C,\mathrm{TM}}= \left[ 0, 0, \cos\left( \pi z /l_{\rm C} \right)\right]$ and $\mathbf{J}_{C,\mathrm{TE}}= \left(0, 1, 0\right)$, respectively. Radiation efficiency is defined as the ratio of the power radiated into free space to the supplied power (integrated at the surface of the capsule). Since the source feed is not taken into account in this capsule-shaped implanted source, the radiation efficiency here does not include mismatch loss that real-world antennas experience. The analysis procedure of this capsule-shaped implanted source model follows the methodology given in \cite{nikolayev2019optimal}.

We first consider a TM or TE source with $r_{\rm impl} = 5$ mm implanted at the center point of a spherical body phantom with the radius $r_{\rm body}=5$ cm. The phantom is homogeneously composed of muscle tissue. For $f \in [0.1, 4]$~GHz, the radiation efficiency of an implanted dipole source (computed using both SWE and the approximation method) is demonstrated as a function of $f$, as shown in Fig.~\ref{fig12}(b). For each type of source, the results obtained by two different methods are almost superimposed. Furthermore, the result of the realistic source model is also plotted, in which the achieved radiation efficiency closely approaches the theoretical bounds of the ideal dipole source model. These results demonstrate that selecting the optimal frequency can effectively increase the achievable radiation efficiency of the implanted antenna even tens of times.

Moreover, the obtained theoretical results (\textit{via} the approximation method) in Fig.~\ref{fig5}(d) and Fig.~\ref{fig8} are validated numerically by the realistic source model, in which the results regarding the optimal frequency or the maximum radiation efficiency are denoted as additional points in corresponding figures. The numerical results from the realistic model are always in close agreement with the theoretical values, which supports the presented analysis and rules on the threshold radius and the maximum achievable radiation efficiency. A supplementary result is the comparison of the curves of $f_{\rm opt}$ for TE sources with different $r_{\rm impl}$, as shown in Fig.~\ref{fig12}(c). Due to relatively limited loss in the reactive near field, only slight deviations can be found among the curves of $f_{\rm opt}$, but similar rules still follow as they do in the TM source case. Similarly, most of the results from the realistic model match well with $f_{\rm opt}$. Only when $r_{\rm body}=$ 2 cm (i.e., $r_{\rm body}$ is within the near field region of the source inside the phantom), deviations could be found due to approximation errors. Similar phenomena have also been found in Fig.~\ref{fig2} when comparing the results of the same model \textit{via} different methods (especially for the TE sources case when $r_{\rm body}$ is relatively small). In fact, if we adopt the acceptable range of $f_{\rm opt}$ with the efficiency tolerance of 0.8, the results from the realistic model still meet the expectations of $f_{\rm opt}$ obtained via the approximation method.

\section*{Authors' contributions}
MG conducted theoretical modeling, computation, results analysis, and numerical verification. DN performed the numerical and experimental verification and result analysis. ZS and AS constructed computational models and directed the overall research. MG and DN drafted the manuscript. All authors read and approved the final manuscript.

\section*{Funding}%% if any
This study was supported in part by the French Agence Nationale de la Recherche (ANR) through the Project MedWave under Grant ANR-21-CE19-0045.

\bibliography{library}

\end{document}